\documentclass[preprint,12pt]{elsarticle}

\usepackage{epsfig}

 \usepackage{graphicx}      % include this line if your document contains figures
\usepackage{amsmath, amssymb}
\usepackage{booktabs}
\usepackage{textcomp} % for Unicode minus sign

\usepackage{algorithm}
\usepackage{algorithmicx} % for \Switch and \Case

\usepackage{algpseudocode}
 \usepackage{multirow} 
 \usepackage{makecell}
\usepackage{subcaption}   % for subfigures
\usepackage{caption}      % optional for better captions
\usepackage{tabularx}

\usepackage{tikz}
\usepackage{tikz-3dplot}
\usetikzlibrary{positioning}
 \usepackage{stfloats}  % in preamble
\usepackage{xcolor}
\algnewcommand{\algorithmicswitch}{\textbf{switch}}
\algnewcommand{\algorithmiccase}{\textbf{case}}
\algnewcommand{\algorithmicendswitch}{\textbf{end switch}}
\algnewcommand{\algorithmicendcase}{\textbf{end case}}
\algdef{SE}[SWITCH]{Switch}{EndSwitch}[1]{\algorithmicswitch\ #1}{\algorithmicendswitch}
\algdef{SE}[CASE]{Case}{EndCase}[1]{\algorithmiccase\ #1}{\algorithmicendcase}
\algtext*{EndSwitch}
\algtext*{EndCase}

\begin{document}
 \begin{frontmatter}
 
\title{Route–Phasing–Split‑Encoded Genetic Algorithm for Multi‑Satellite On‑Orbit Servicing Mission Planning}

\author{Shridhar  Velhal, Avijit Banerjee, and George Nikolakopoulos}

\affiliation{organization={Robotics and AI Group, Luleå University of Technology},%Department and Organization
            city={Luleå},            
            country={Sweden}
            }

\begin{abstract}
This article addresses multi-servicer on-orbit servicing mission planning in geosynchronous Earth orbit, where routing decisions are tightly coupled with time-dependent orbital phasing and strict propellant and mission-duration constraints. We propose a Route–Phasing–Split Genetic Algorithm (RPS-GA) that simultaneously optimizes target sequencing, discrete phasing rotation decisions (i.e., the number of phasing revolutions/waiting cycles), and route partitioning across multiple servicing spacecrafts (SSCs). An RPS triplet chromosome encodes route order, phasing rotations, and route splits in a unified structure, enabling split-aware recombination without disrupting feasible multi-servicer route blocks. 
Feasibility is enforced through a constraint-aware fitness function that ranks feasible solutions based on total $\Delta V$, while penalizing propellant and mission duration violations, using aggregate and imbalance penalties. This formulation discourages the concentration of violations on a single servicing spacecraft (SSC). Once a feasible best solution is identified, it is preserved as feasible in subsequent generations, thereby enhancing convergence stability.
The framework incorporates split-aware crossover, mutation and a regret-based Large Neighborhood Search for local intensification. Experiments on representative GEO servicing scenarios demonstrate that RPS-GA produces feasible multi-servicer plans with substantially improved fuel efficiency, reducing total $\Delta V$ by $24.5\%$, (from $1956.36 \  m/s$   to $ 1476.32\ m/s $) compared with a state-of-the-art LNS-AGA baseline.
\end{abstract}

%\item Integrated optimization of routing, phasing, and task allocation
%\item Route–Phasing–Split chromosome for GEO servicing
%\item Feasibility-preserving fitness for multi-SSC missions
%\item 24.5\%  reduction in total $\Delta V$  over  state-of-the-art LNS-AGA in benchmark case.

\end{frontmatter}

\section{Introduction}
With the increasing emphasis on sustainable space operations, enhanced on-orbit servicing (OOS) is emerging as a foundational capability for future space infrastructure, supporting satellite life extension, constellation maintenance, failure recovery, and long-term orbital sustainability \cite{banerjee2023resiliency}. As demand for such services increases, efficient orbital routing in geosynchronous Earth orbit (GEO) becomes critical, given the high value and long life-span of GEO satellites~\cite{barbara2020new,liu2021economic}. As demand for such services grows, Servicing spacecrafts (SSCs) must transfer sequentially among multiple targets, rendering route sequencing and timing central for ensuring mission feasibility and propellant efficiency. In GEO, maneuver cost is dominated by total $\Delta V$, while feasibility is strongly constrained by orbital phasing. Even small inefficiencies in visit order or departure timing can result in significant propellant penalties and reduced operational lifetime. 

GEO orbital routing poses additional challenges due to the continuous evolution of relative satellite geometry, which renders transfer feasibility and cost inherently time-dependent, and they are summarized in a \cite{opromolla2024future}. From a planning perspective, on-orbit routing can be viewed as a spatio-temporal multi-task assignment (STMTA) problem, since it must decide both which targets to service and when to execute transfers (i.e., from where and when to maneuver) under strict timing constraints. However, unlike standard STMTA formulations \cite{velhal2022decentralized,velhal2022dynamic} that typically assume stationary travel costs between tasks, orbital transfers induce nonstationary transition costs of the ($\Delta V$) and even feasibility of moving between the same pair of targets vary  due to phasing. Planning therefore needs to balance the objective of minimizing ($\Delta V$) while satisfying tight time windows and maintaining feasible phasing, all under the strong coupling between sequence, split, and orbital synchronization decisions. Large constellations and multi-servicer missions further demand scalable, near-real-time solutions, which traditional deterministic methods often cannot provide as problem size and constraint complexity increase. These factors motivate adaptive optimization frameworks capable of efficiently exploring large combinatorial spaces while preserving mission feasibility under nonlinear orbital dynamics.

Within on-orbit servicing, routing formulations vary with mission objectives and task heterogeneity. Refueling missions \cite{zhou2017multi,zhang2019optimal,kong2023multiobjective} primarily involve transferring propellant from servicing spacecraft or depots to fuel-deficient targets, creating capacity and logistics constrained routing problems where rendezvous and return maneuvers dominate cost. Repair or maintenance missions \cite{han2022multiple} introduce variable service durations, urgency constraints, and specialized servicing capabilities, often requiring strict deadlines. Mixed missions \cite{liang2024geo, hudson2020versatile} combine multiple service types and may involve uncertainty in service success or duration, increasing task heterogeneity and the complexity of feasible routing. Recently, \cite{yan2024rapid} explicitly distinguishes between full-service and quick-service policies to maximize task completion under tight propellant and time budgets.  This heterogeneity directly impacts routing representations and optimization strategies, as it introduces strong coupling between route sequencing, split/allocation decisions, and orbital phasing. Route feasibility and transfer cost are typically estimated using impulsive, near-circular maneuvers such as Hohmann-type transfers or Lambert targeting \cite{xie2018hohmann, daneshjou2017mission}, highlighting the need for flexible, mission-feasible solution encodings.

\begin{figure}[htbp]
\begin{subfigure}[t]{0.98\linewidth}
\centering
\includegraphics[width=\linewidth,  trim=120 120 100 100, clip]{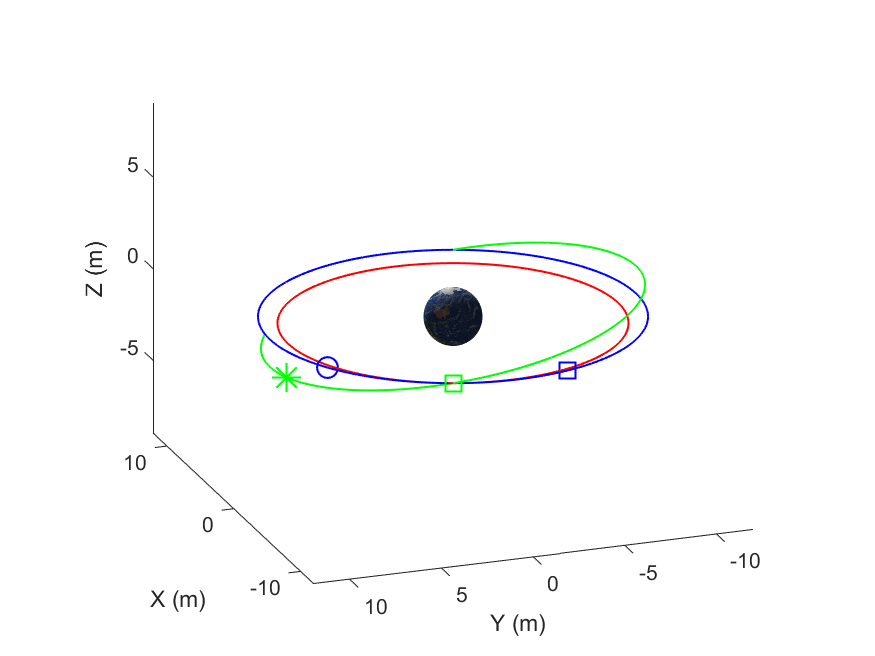} \\
\caption{Orbital transfer incorporating change of plane as well as phasing, a representative scenario transferring  from green orbit to blue orbit. $\ast$  denotes the initial orbital position of service satellites, $\circ$  denote the initial positions of target satellites , $\square $ denotes the position at the plane change transfer. After phasing revolutions on the red orbit, SCC rendezvous with the target satellite.}
\label{fig:one_transfer}
\end{subfigure} \\
\begin{subfigure}[t]{0.98\linewidth}
\centering
\includegraphics[width=\linewidth, trim=120 80 80 80, clip]{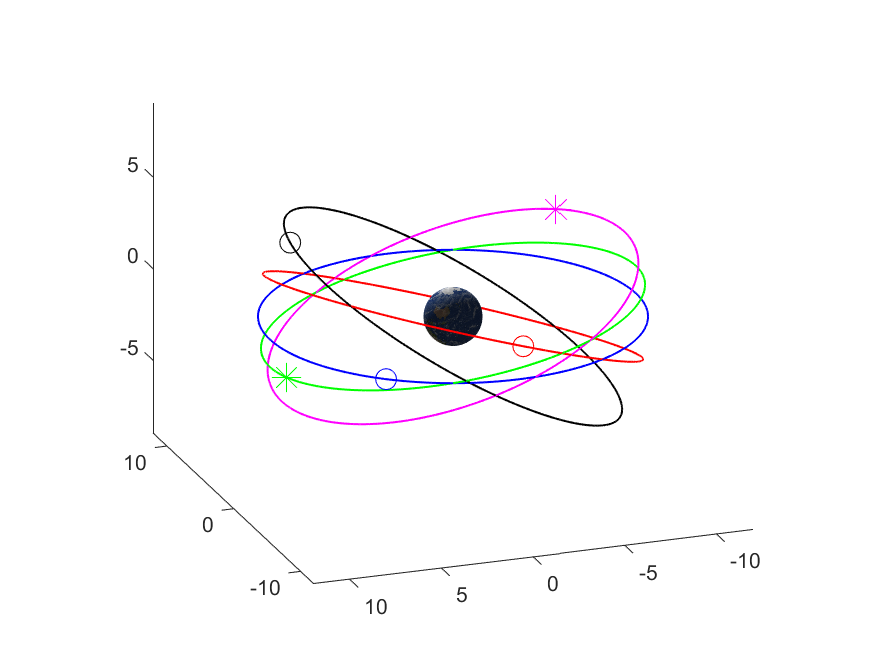} \\
\caption{  SSCs need to find the sequence in which they service all satellites within  $\Delta V$ and mission deadline constraints.  $\ast$  denotes the initial orbital position of service satellites, $\circ$  denote the initial positions of target satellites  }
\label{fig:phasing_type_0}
\end{subfigure} 
\caption{A snapshot of on-Orbit servicing mission  } 
\label{fig:concept}
\end{figure}

Figure \ref{fig:concept} shows the on orbit routing problem in which SSC needs to find the optimal sequence for visiting target orbits. The SSC needs to coast until the intersection point of source and target orbits, then it needs to perform phasing to rendezvous at the target satellite. A central modeling difficulty in GEO routing is that transfer feasibility and $\Delta V$ cost depend on phasing. Unlike terrestrial routing, the “travel cost” between two targets is non-stationary.  For a plane change, the SSC needs to wait for a favorable alignment, and for phasing corrections, it requires several orbital revolutions to synchronize the transfer. Consequently, sequencing alone is insufficient; effective planning must jointly determine the route of SSCs, phasing rotations (i.e., phasing-time and cost), and how targets are partitioned and assigned across multiple servicing spacecraft. This tight coupling distinguishes orbital routing from classical vehicle routing problems (VRPs) and motivates specialized representations and search operators to maintain feasibility and avoid repeated violations during optimization.

Earlier studies have progressively adapted classical routing formulations to reflect orbital constraints. Single-servicer GEO missions have been modeled as time-dependent TSP-like problems using transfer costs derived from Hohmann or Lambert-based rendezvous \cite{zhou2017multi, xie2018hohmann, daneshjou2017mission}. Deadline and window-constrained variants introduce VRPTW structure, where accessibility windows and mission completion constraints govern feasibility \cite{han2022multiple}.
To handle larger instances, clustered formulations partition targets by orbital proximity or demand and then coordinate intra/inter cluster routing \cite{yin2025clustering}. Multi-servicer and depot-enabled scenarios yield multi-depot VRP variants in which servicing spacecraft shuttle between depots and targets under capacity and timing constraints \cite{kong2023multiobjective}, while service-station architectures and multi-orbit servicing generalize these models to hub-and-spoke or network logistics structures \cite{qiao2024optimal, sorenson2023multi}. These developments reflect increasing recognition that orbital routing must simultaneously capture time dependence, capacity logistics, and multi-agent coordination.

Solution techniques range from exact optimization to heuristics and learning-based replanning, with scalability as the primary differentiator. Exact methods such as MILP and time-expanded network formulations are effective for strategic analysis and small instances, including architecture evaluation with depots and multi-commodity flows \cite{sarton2021framework, sorenson2023multi}. However, the strong coupling between discrete routing decisions and time-dependent phasing typically limits exact approaches as the number of targets or SSCs increases. Metaheuristics are therefore prevalent in operational settings. Genetic Algorithms (GA) and Non-dominated Sorting Genetic Algorithm II (NSGA-II) are widely used for sequencing and multi-objective tradeoffs in refueling and repair planning \cite{han2022multiple, zhou2017multi, kong2023multiobjective}, while Particle Swarm Optimization (PSO)-based methods are employed when continuous variables such as departure times or transfer durations must be optimized alongside assignments \cite{li2021deployment, zhou2015mission, daneshjou2017mission}. Large Neighborhood Search (LNS) and Adaptive Large Neighborhood Search (ALNS) frameworks provide robust feasibility handling through destroy-and-repair cycles with adaptive operator selection \cite{ropke2006adaptive, chen2020adaptive}, and hybrid GA–LNS methods have demonstrated improved performance in tightly constrained GEO scenarios \cite{han2022multiple, yan2024rapid}. For large-scale settings, clustering–reconstruction and hierarchical decomposition are commonly used to reduce problem dimension and coordinate multi-SSc decisions \cite{yin2025clustering, wu2023hierarchical, li2025multi}. More recently, learning-based methods such as reinforcement learning \cite{wu2025geo} and pointer-network approaches \cite{tomanek2025pointer} have been explored for rapid sequencing and replanning, trading strict optimality guarantees for speed and adaptability, and Monte Carlo Tree Search (MCTS)–based approaches \cite{ye2025improved} have also been proposed for multi-spacecraft allocation.

Despite these advancements, multi-servicer GEO routing with time~–~dependent feasibility still faces three persistent challenges.
(i) feasibility is often enforced through penalty and repair steps in a way that allows constraint violations to oscillate among servicing spacecraft, rather than converging to a jointly feasible plan. 
(ii) Routing, phasing time, and multi-servicer partitioning are often optimized sequentially or weakly coupled. However, in on-orbit routing these decisions are tightly linked; changing the visit order can invalidate feasible phasing windows and transfer timing, while reassigning targets can disrupt route partitions. Optimizing these layers independently can therefore induce repeated feasibility violations during search, as improvements in one dimension inadvertently degrade another.
(iii) Many genetic and neighborhood operators used in orbital routing are adapted from terrestrial VRP methods designed for stationary edge costs. While they preserve adjacency or partition structure, recombination can disrupt route blocks and phasing/waiting sequences, leading to violations of time windows and propellant constraints under time-dependent transfer costs.
Consequently, multi-servicer orbital routing is an NP-hard problem combining VRP-like combinatorial complexity with nonlinear, time-dependent orbital constraints. Effective methods must explore large discrete spaces while preserving feasibility and minimizing repeated violations during search.

To address these challenges, this paper proposes a Route–Phasing–Split Genetic Algorithm (RPS-GA) for on-orbit servicing mission planning. The method introduces an RPS triplet chromosome that jointly encodes the route sequence, phasing rotations, and route splits in a single integrated structure, enabling integrated optimization of the coupled decisions under tight propellant and mission-duration constraints. In addition, we propose a constraint-aware fitness to manage feasibility reliably. Feasible solutions are compared only by total $\Delta V$ while infeasible solutions are penalized both for aggregate constraint violations across all servicing spacecraft and for unbalanced violations in which a single spacecraft carries the majority of infeasibility. This design encourages convergence toward globally feasible multi-servicer plans rather than merely shifting infeasibility between routes. Moreover, the penalty structure ensures forward invariance of the best-found solution i.e.,  once a feasible solution is found, the best chromosome remains feasible in subsequent generations, preventing infeasible low-($\Delta V$) solutions from being selected as best

The framework is further enhanced with operators tailored to the Route – Phasing-Split (RPS) representation. Split-aware genetic operators—Route-Block Crossover (RBC) and adaptive mutation, explicitly preserve and recombine feasible route blocks, reducing disruption of high-quality partitions during reproduction. To strengthen local improvement, a regret-based Large Neighborhood Search is embedded using 2-regret insertion and rotation-neighborhood exploration to refine sequences and exploit favorable phasing opportunities. Experimental results on the GEO servicing scenario show that RPS-GA substantially improves the fuel efficiency under identical evaluation settings, reducing total $\Delta V$ by $24.5\% $,  (from $ 1956.36 m/s$   to $ 1476.32  m/s $) compared with state-of-the-art baselines\cite{han2022multiple}, while maintaining feasibility for multi-servicer mission plans.

The main contributions of this paper are as follows:
\begin{enumerate}
\item Novel RPS Triplet Chromosome: A Route–Phasing–Split (RPS) representation that encodes route sequences, phasing rotations, and route splits in a single structure, enabling efficient exploration of complex orbital routing solutions under tight propellant and mission deadline constraints.
\item Constraint-Aware Fitness with Forward-Invariant Best Solution: A penalty mechanism that preserves the feasibility of the best-found solution across generations, guiding the search toward physically realizable multi-servicer plans while reducing the tendency to shift constraint violations between routes.
\item Significant $\Delta V$  Reduction: The proposed method achieves a significant improvement over state-of-the-art LNS-AGA \cite{han2022multiple} approach, reducing total $\Delta V$ from $1956.36 m/s$ to $1476.32 m/s$, demonstrating enhanced fuel efficiency and optimized orbital maneuver planning.
 \end{enumerate}

The rest of the paper is organized as follows. Section~\ref{sec:orbital_transfers} reviews the on-orbit transfer model, including transfer time and ($\Delta V$) computation. Section~\ref{sec:problem_formulation} formulates the multi-servicer GEO routing problem and its constraints. Section~\ref{sec:ga_methodology} presents RPS-GA, including the RPS triplet chromosome, constraint-aware fitness, and split-aware operators. Section~\ref{sec:results} reports the experimental setup and comparative results. Section~\ref{sec:conclusion} concludes the paper.

% PRELIMINARIES ON ORBITAL TRANSFERS
\section{Preliminaries on Orbital Transfers}
\label{sec:orbital_transfers}

The on-orbit scheduling problem aims to plan transfers of multiple satellites between different orbital tasks while minimizing total propellant consumption, quantified by the $\Delta V$ cost. Each satellite starts from a distinct initial orbit and may perform maneuvers such as plane changes and phasing adjustments to reach the target orbit. This section details the orbital transfer strategy for servicing missions between geosynchronous satellites.  

\subsection{Transfer Method Overview}

The transfer method consists of two main maneuvers:

\begin{enumerate}
    \item \textbf{Plane-Change Maneuver:} Adjust the inclination and orientation of the orbit to match the target orbital plane.
    \item \textbf{Phasing Maneuver:} Modify the orbit temporarily (in the  target plane) to synchronize the satellite with the target phase.
\end{enumerate}
The total $\Delta V$ cost of a transfer is the sum of the impulses required for these maneuvers. Intermediate points along the orbital plane intersection are selected to minimize coast times, and velocity vectors are computed at these maneuver points to determine the required impulses.

\subsection{Mathematical Formulation of $\Delta V$ and Temporal Components}

Let the source and target orbits be characterized by semi-major axes $a_s$, $a_t$, inclinations $i_s$, $i_t$, and right ascension of ascending nodes (RAANs) $\Omega_s$, $\Omega_t$. The gravitational parameter is denoted $\mu$.

\paragraph{Angular Separation Between Orbits:}  The angular separation can be calculated as
\begin{equation}
    \alpha = \arccos\Big( \cos i_s \cos i_t + \sin i_s \sin i_t \cos(\Omega_s - \Omega_t) \Big)
\end{equation}

\subsubsection{Plane Change Maneuver} 
The plane change maneuver aligns the SSC (Service Spacecraft) with the target's orbital plane. This maneuver can only occur at the intersection points of the two orbital planes, providing two opportunities per orbital period.

The intersection points are:
\begin{align}
\mathbf{r}_{m1} = a \frac{\mathbf{h}_s \times \mathbf{h}_t}{\|\mathbf{h}_s \times \mathbf{h}_t\|}, \quad
\mathbf{r}_{m2} = -\mathbf{r}_{m1}
\label{eq:intersection_points}
\end{align}
where $a$ is the GEO orbit radius, and $\mathbf{h}_s$, $\mathbf{h}_t$ are angular momentum vectors. The angular momentum vectors can be computed as
\begin{align*}
\mathbf{h} &= \mathbf{r} \times \mathbf{v} \\
  &= \sqrt{\mu a} 
\begin{bmatrix}
\cos\Omega & -\sin\Omega & 0 \\
\sin\Omega & \cos\Omega & 0 \\
0 & 0 & 1
\end{bmatrix}
\begin{bmatrix}
1 & 0 & 0 \\
0 & \cos i & -\sin i \\
0 & \sin i & \cos i
\end{bmatrix}
\begin{bmatrix}
0 \\ 0 \\ 1
\end{bmatrix}
\label{eq:angular_momentum}
\end{align*}
where $\mu$ is Earth's gravitational constant.

\paragraph{$ {\Delta v}_{plane}$  } 
The plane change impulse at the intersection point is:
\begin{align}
\mathbf{\Delta v}_{plane} &= \mathbf{v}_t - \mathbf{v}_s, \\
\|\mathbf{\Delta v}_{plane} \| &= 2\|\mathbf{v}_t\|\sin(\alpha/2)
\label{eq:plane_change_dv}
\end{align}
where $\mathbf{v}_s$ and $\mathbf{v}_t$ are velocities at the maneuver point, and $\mathbf{v}$ is the circular orbit velocity magnitude.

\paragraph{Coast Time ($t_\text{coast}$):} 
Time to travel along the orbit to the plane-change maneuver point:
\begin{align}
t_\text{coast} &= \frac{\Delta \theta}{\omega_\text{GEO}}  \nonumber \\
t_{\text{coast}} &=
\begin{cases}
 \dfrac{\arccos\left(\dfrac{\mathbf{r}_{m1} \cdot \mathbf{r}_0}{\|\mathbf{r}_{m1}\|\|\mathbf{r}_0\|}\right)}{\omega_\text{GEO}}, & (\mathbf{r}_0 \times \mathbf{r}_{m1}) \cdot \mathbf{h}_s > 0 \\[12pt]
 \dfrac{\pi - \arccos\left(\dfrac{\mathbf{r}_{m1} \cdot \mathbf{r}_0}{\|\mathbf{r}_{m1}\|\|\mathbf{r}_0\|}\right) }{\omega_\text{GEO}}, & (\mathbf{r}_0 \times \mathbf{r}_{m1}) \cdot \mathbf{h}_s < 0
\end{cases}
\label{eq:coast_time}
\end{align}

GEO satellites maintain a circular orbit at $a_\text{GEO} \approx 42,164~\text{km}$, with orbital period of 24 hours. All satellites share the same angular velocity $\omega_\text{GEO}$, and relative positions remain fixed in the Earth-centered rotating frame. For circular GEO orbit, $t_\text{coast}$ is constant and independent of mission start time.

\subsubsection{Phasing Maneuver}
After plane alignment, the SSC must synchronize its position with the target through a phasing maneuver. The phasing angle $\theta$ between SSC and target is:
\begin{align}
\theta =
\begin{cases}
\psi, & |\psi| \leq \pi \\
-2\pi + |\psi|, & \psi > \pi \\
2\pi - |\psi|, & \psi < -\pi
\end{cases}
\label{eq:phasing_angle}
\end{align}
where $\psi = (\Omega_s + \omega_s) - (\Omega_t + \omega_t)$, with $\omega_s, \omega_t$ being true anomalies.

\paragraph{Phasing Revolutions}
The SSC enters a phasing orbit tangent to the original orbit, completing $k_{\text{phase}}$ revolutions while the target completes $k_{\text{GEO}}$ revolutions. The phasing time is:
\begin{align}
t_{\text{phase}} = \frac{2\pi k_{\text{GEO}} + \theta}{2\pi} T_{\text{GEO}}
\label{eq:phasing_time}
\end{align}

The corresponding semi-major axis of the phasing orbit is:
\begin{align}
a_{\text{phase}} = r_{\text{GEO}} \left( \frac{2\pi k_{\text{GEO}} + \theta}{2\pi k_{\text{GEO}}} \right)^{2/3}
\label{eq:phasing_orbit}
\end{align}

\paragraph{Delta-V for Phasing}
The total phasing delta-v (split into two equal and opposite impulses) is:
\begin{align}
\Delta v_{\text{phase}} = 2 \left| \sqrt{\mu} \left( \sqrt{\frac{2}{r_{\text{GEO}}} - \frac{1}{a_{\text{phase}}}} - \sqrt{\frac{1}{r_{\text{GEO}}}} \right) \right|
\label{eq:phasing_dv}
\end{align}

The individual impulses aligned with the velocity direction $\hat{\mathbf{v}}_m$ are:
\begin{align}
\mathbf{\Delta v}_{phase1} &= \frac{1}{2} \Delta v_{\text{phase}} \operatorname{sgn}(\theta) \frac{\mathbf{v}_m}{\|\mathbf{v}_m\|}, \\
\mathbf{\Delta v}_{phase2} &= -\mathbf{\Delta v}_{phase1}
\label{eq:phasing_impulses}
\end{align}
where $\mathbf{v}_m$ is the target velocity at the maneuver point.

\subsubsection{Combined Plane Change and Phasing Maneuver Strategy}

\begin{figure}[htbp]
\centering
\begin{subfigure}[t]{0.48\linewidth}
\centering
\includegraphics[width=\linewidth]{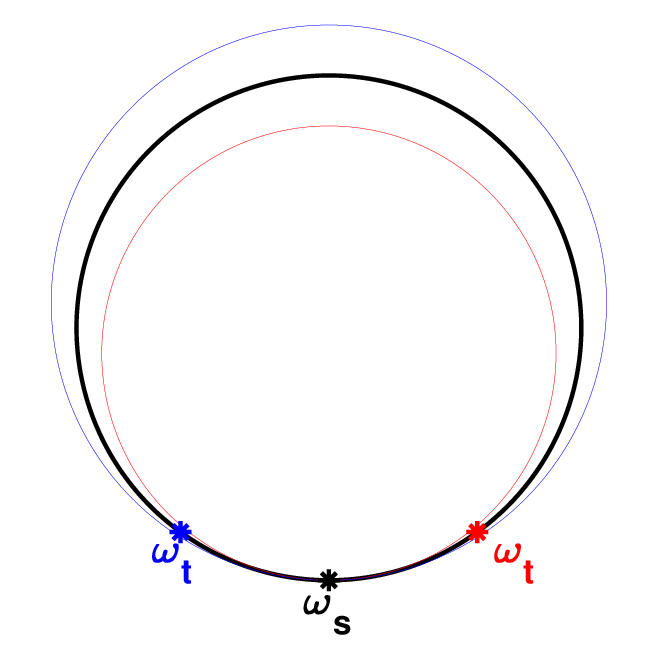}
\caption{Phasing maneuvers for positive and negative angles}
\label{fig:phasing_type}
\end{subfigure}
\hfill
\begin{subfigure}[t]{0.48\linewidth}
\centering
\includegraphics[width=\linewidth]{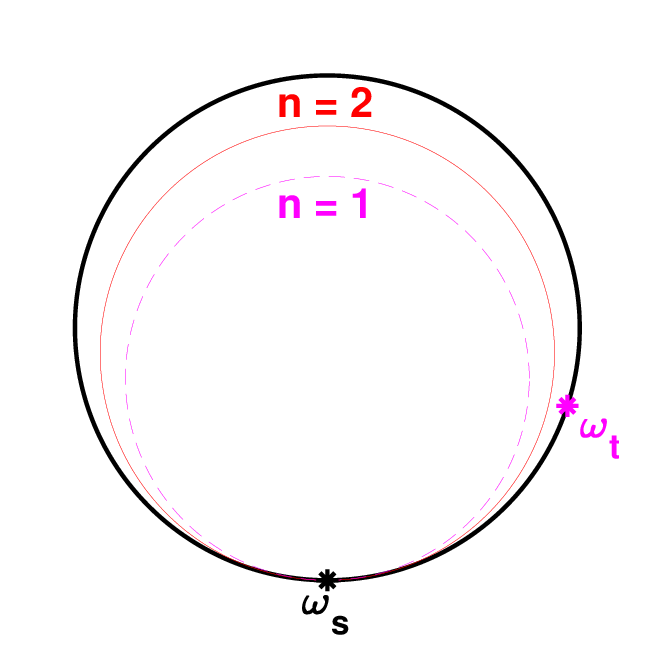}
\caption{Multi-rotation phasing}
\label{fig:combined_maneuver}
\end{subfigure}
\caption{Phasing maneuvers }
\label{fig:phasing_combined}
\end{figure}

The plane change and phasing maneuvers can be combined by executing both at the intersection point. As shown in Figure \ref{fig:combined_maneuver}, $\mathbf{\Delta v}_1$ and $\mathbf{\Delta v}_2$ merge into a single impulse, reducing the total delta-v cost. The second impulse will be applied using in-plane impuelse after achieving the desired phase angle.

The first impulse is computed as 
\begin{align}
\mathbf{\Delta v}_{\text{1}} &= \|\mathbf{\Delta v}_{plane} + \mathbf{\Delta v}_{phase1}\|  \label{eq:deltaV_combined01}  
%t &= t_{coast}
\end{align}

The second impulse is  
\begin{align}
\mathbf{\Delta v}_{\text{2}}  &= \mathbf{\Delta v}_{phase2} \label{eq:deltaV_combined02}  
%t &= t_{phase}
\end{align}

The total efforts are
\begin{align}
 {\Delta v}_{\text{total}} &= \|\mathbf{\Delta v}_{plane} + \mathbf{\Delta v}_{phase1}\|  + \| \mathbf{\Delta v}_{\text{2}} \| \label{eq:deltaV_combined} \\
t &= t_{coast} +  t_{phase} \label{eq:total_time} 
\end{align}

\begin{figure}[htbp]
\centering
\includegraphics[width=\linewidth, trim=60 240 60 290, clip]{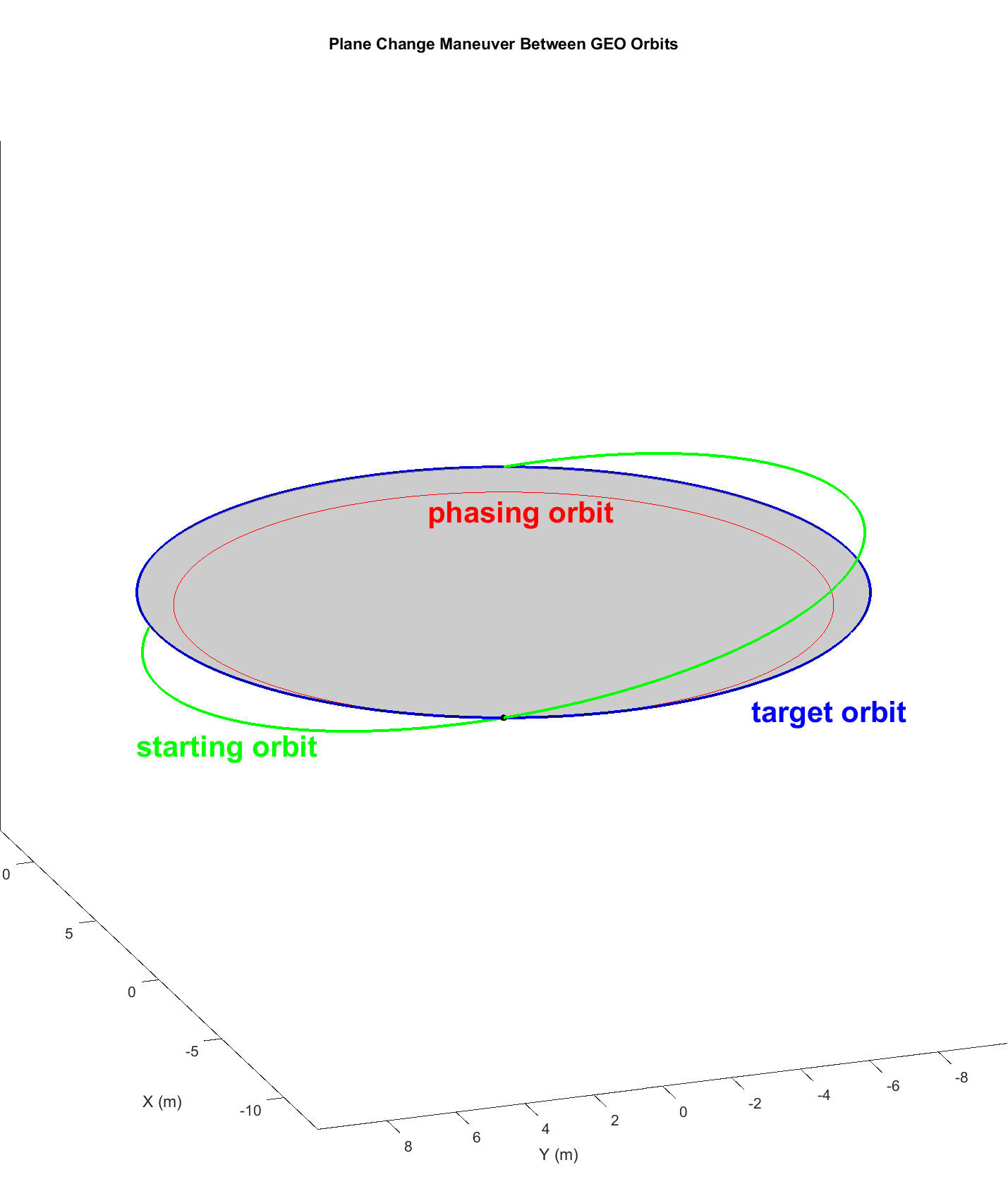}
\includegraphics[width=0.8\linewidth, trim=0 0 0 20, clip]{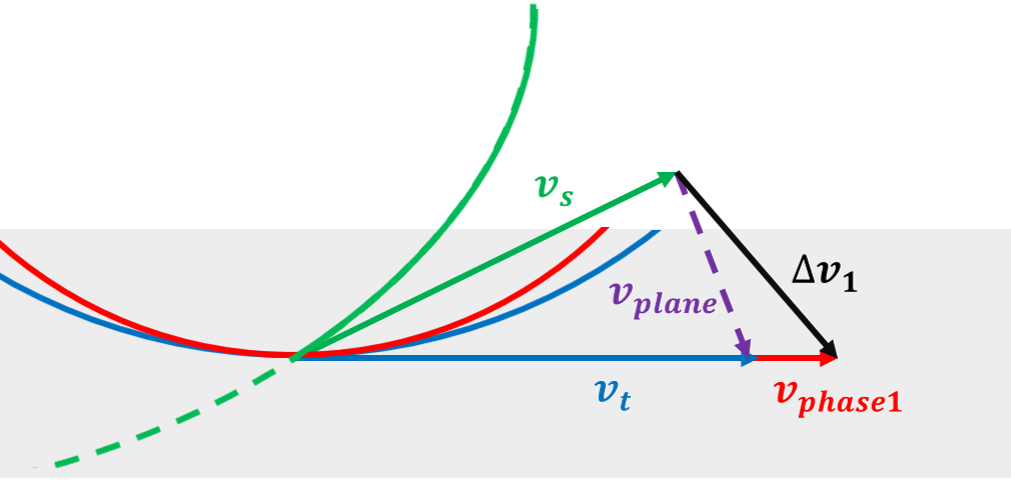}
\caption{Combined plane-change and phasing maneuver illustrating $\Delta V $ components}
\label{fig:Plane_change}
\end{figure}

\subsubsection{Waiting Time }   
The phase change maneuver starts from either of two intersection points $(\mathbf{r}_{m1} $  and $\mathbf{r}_{m2}$).  After reaching the targeted orbit, SSC repairs the satellite. After this repair task, SSC needs to stay in the same orbit till the closest intersection point for the next maneuver. This time is referred to as a waiting time. 
For GEO-to-GEO transfers, coast and phasing times are constants, only $t_\text{wait}$ may vary. It varies with the sequence in which the SSC travels through the orbits.

\subsection{Summary}

This orbital transfer strategy combines plane change and phasing maneuvers into an efficient two-impulse sequence. The total delta-v (Eq. \ref{eq:deltaV_combined}) and transfer time (Eq. \ref{eq:total_time}) are functions of the phasing revolution count $k_{\text{GEO}}$, which serves as an integer decision variable in the mission planning optimization.

\section{Multi-TSP Formulation for Orbital Routing in GEO}
\label{sec:problem_formulation}
Let $m$ represent the total number of servicing spacecraft (SSC), and $\mathcal{S} = \{1,2,\dots,m\}$  denotes the index set of SSCs. 
Let $n_t$ represent the number of repair tasks in the mission.  The index set of repair tasks is denoted as  $\mathcal{T} = \{1,2,\dots,n_t\}$. In general $n_t >> m$. 

The on-orbit repairing problem considers the multi satellite orbital routing problem in which each SSCs, operating in geosynchronous orbit, sequentially visits multiple geosynchronous orbital satellites to repair them. The route of $S_i$ , denoted as $\mu_i$, is the ordered sequence of repair tasks to be executed.
The objective is to determine the routes and maneuver parameters that minimize the total orbital transfer cost, measured in terms of cumulative $\Delta V$ required by the team of SSCs for the entire mission.

In the proposed on-orbit servicing framework, each SSC is assigned
a mission route that specifies the sequence of GEO satellites to be serviced.
The mathematical characteristics of these routes are defined as:
\begin{enumerate}
    \item route $\mu_i $   consists only of valid GEO servicing tasks  $ \Rightarrow \mu_i \subseteq \mathcal{T}$ 
    \item Each GEO satellite is visited at most once by SSC $ \Rightarrow \forall j,k \in \mu_i,\; j \neq k,$
    \item Each servicing task is assigned to exactly one agent $ \Rightarrow \mu_i \cap \mu_l = \varnothing, \quad \forall i \neq l, $
\end{enumerate}

These conditions ensure that each servicing spacecraft follows a physically meaningful and non-redundant orbital route, while preventing conflicting task assignments among multiple agents. 
In the GEO servicing context, this guarantees unique assignment of each target satellite, avoids duplicate rendezvous maneuvers, and enables efficient fuel usage for plane-change and phasing transfers.

\subsection{Route Continuity and Assignment Variables}
For each servicing spacecraft (SSC) $S_i$ with assigned route $\mu_i$, task assignment and route connectivity are captured by two sets of binary variables:

Task assignment variable,  $ x_{ij} \in \{0,1\} $  indicates whether SSC $S_i$ is assigned for servicing task $T_j$ or not?. The transition variable $ y_{kj}^i \in \{0,1\}  $.  
$y_{kj}^i$ is a binary variable equal to one if SSC $S_i$ directly transits from orbit $k$ to execute task $T_j$ in its route. This focuses on the sequence of tasks in path.

Each task must be serviced by exactly one SSC, ensuring complete coverage without duplication.
\begin{equation} 
\sum_{i \in \mathcal{S} } x_{ij} = 1,  \quad \forall j \in \mathcal{T}, \label{eq:conntinuity_1}  
\end{equation} 

Each task in a route must have exactly one incoming edge (either from a previous task or the start node), conditional on being assigned to SSC $i$.
  \begin{equation} 
  \sum_{k \in \{ \mathcal{T} \cup \mathcal{S} \} } y_{kj}^i = x_{ij},  \quad
 \forall i \in \mathcal{S}, \forall j \in \mathcal{T}, \label{eq:conntinuity_2}  
 \end{equation}

Each task and SSC in a route must have exactly one outgoing edge to the next task, conditional on being assigned to SSC $i$.
\begin{equation} 
  \sum_{j \in \mathcal{T} } y_{kj}^i = x_{ij}, \quad \forall i \in \mathcal{S}, \forall k \in \{ \mathcal{T} \cup \mathcal{S} \}, \label{eq:conntinuity_3}  
 \end{equation} 
 
Self-transitions are prohibited to avoid trivial loops.
  \begin{equation} 
  y_{jj}^i = 0, \quad \forall i \in \mathcal{S}, \forall j \in \mathcal{T}. \label{eq:conntinuity_4}  
 \end{equation}

\subsection{Cost Function}
The objective of the mission is to execute all the repair tasks with minimum  total $\Delta V$.  
Unlike classical TSP formulations, the cost of traversing an edge depends not only on the task sequence but also on the number of phasing orbits executed between successive tasks.
Let $ n_{kj}^i \in \mathbb{Z}^+ $ ,
denote the number of phasing revolutions performed by   $S_i$ when transitioning from
task $T_k$ to task $T_j$.

The maneuver cost for  $S_i$ transitioning from task $T_k$ to task $T_j$ is then given by
\begin{equation}
\Delta V_{kj}^i = f(k,j,n_{kj}^i),
\end{equation}
 
where $f(\cdot)$ captures the $\Delta V$ required for the Hohmann transfer and associated
phasing adjustments in GEO and computed using eq \eqref{eq:deltaV_combined}. $\Delta V_{kj}^i$ is a function of task $T_k$ and $T_j$, and this does not depend on the remaining tasks in route.

The total cost of route $\mu_i$ is defined as the sum of the costs of all transitions in a route, 
\begin{equation}
\Delta V_i(\mu_i) = \sum_{(k,j) \in \mu_i} \Delta V_{kj}^i,
\end{equation}
and the overall mission cost is the sum over all SSCs.

\subsection{Temporal Constraints and Mission Completion}
\label{subsec:temporal_constraints}
Each orbital transfer and servicing operation incurs not only energy i.e.  a $\Delta V$ cost but also a
temporal cost, which depends on the selected phasing maneuver. 

Let us define the notation used for temporal variables:
\begin{itemize}
    \item \( t_{\text{arrive}}^{ik} \): Arrival time of spacecraft \( i \) at task \( k \)
    \item \( t_{\text{repair}}^{k} \): Repair duration for task \( k \)
    \item \( t_{\text{end}}^{ik} \): Completion time of task \( k \) by spacecraft \( i \)
    %\item \( t_{\text{wait}}^{ikj} \): Waiting time before spacecraft \( i \) initiates transfer from task \( k \) to task \( j \)
    %\item \( t_{\text{transition}}^{ikj} \): Time when spacecraft \( i \) begins the orbital transfer from task \( k \) to task \( j \)
    \item \( t_{\text{coast}}^{ikj} \): Coast time required for spacecraft \( i \) to transfer from task \( k \) to task \( j \)
    \item \( t_{\text{phase}}^{ikj} \): Phasing time required for spacecraft \( i \) to transfer from task \( k \) to task \( j \)
    \item \( t_{\text{max}} \): Mission deadline
\end{itemize}

{\renewcommand{\arraystretch}{1.2}
\begin{table}[h]
\centering
\caption{Temporal parameters and decision variables.}
\label{tab:time_params}
\begin{tabularx}{\linewidth}{lX}
\hline
  \( t_{\text{arrive}}^{ik} \) & Arrival time of spacecraft \( i \) at task \( k \) \\
 \( t_{\text{repair}}^{k} \)  &  Repair duration for task \( k \) \\
 \( t_{\text{end}}^{ik} \) &  Completion time of task \( k \) by spacecraft \( i \) \\
    % \( t_{\text{wait}}^{ikj} \) &  Waiting time before spacecraft \( i \) initiates transfer from task \( k \) to task \( j \) \\
    %  \( t_{\text{transition}}^{ikj} \) & Time when spacecraft \( i \) begins the orbital transfer from task \( k \) to task \( j \) \\
     \( t_{\text{coast}}^{ikj} \) &  Coast time required for spacecraft \( i \) to transfer from task \( k \) to task \( j \) \\
      \( t_{\text{phase}}^{ikj} \)  & Phasing time required for spacecraft \( i \) to transfer from task \( k \) to task \( j \) \\
      \( t_{\text{max}} \)  &  Mission deadline \\
\hline
\end{tabularx}
\end{table}}

The temporal sequence for an orbital transfer executed by spacecraft \( i \) from task \( k \) to task \( j \) consists of five consecutive phases. First, the spacecraft arrives at task \( k \) at time \( t_{\text{arrive}}^{ik} \) and performs the repair operation, which requires a fixed duration \( t_{\text{repair}}^{k} \). Thus the repair is completed at
\begin{align}
t_{\text{end}}^{ik} = t_{\text{arrive}}^{ik} + t_{\text{repair}}^{k}.
\label{eq:repair_completion}
\end{align}

The orbital transfer comprises a coast phase of duration \( t_{\text{coast}}^{ikj} \) followed by a phasing maneuver of duration \( t_{\text{phase}}^{ikj} \), both determined by the orbital mechanics described in Section~\ref{sec:orbital_transfers}. Consequently, the arrival time at the next task \( j \) is
\begin{align}
t_{\text{arrive}}^{ij} = t_{\text{end}}^{ik} + t_{\text{coast}}^{ikj} + t_{\text{phase}}^{ikj}.
\label{eq:arrival_time}
\end{align}

Let $\tau_{kj}^i  $ denote the time required by   $S_i$ to transfer from task $T_k$ to task $T_j$.
This temporal cost is the time required by $S_i$ after the completion of task $T_k$ until it completes the task $T_k$. 

The temporal maneuver cost for  $S_i$ transitioning from task $T_k$ to task $T_j$ is the time required for  coast time, phasing time and the repair time. 
\begin{equation} 
\tau_{kj}^i =    t_{\text{coast}}^{ikj} + t_{\text{phase}}^{ikj} +  t_{\text{repair}}^{j}  \end{equation}

Unlike $\Delta V_{kj}^i$, the $\tau_{kj}^i$ is route-dependent. In geosynchronous orbits, the phasing time depends only on the two orbits involved and the number of phasing revolutions. Also the repair time is fixed for each task. But the coast time depends on position of the spacecraft, i.e., it depends on the start time of task $T_k$. Hence,   $\tau_{kj}^i $ is a function of entire route $\mu_i$ and cannot be computed locally for pair $T_k$ and $T_j$. 

To ensure temporal feasibility in the mission schedule, a Big-M constraint enforces the proper sequencing whenever spacecraft \( i \) travels directly from task \( k \) to task \( j \) (i.e., when \( x_{ikj} = 1 \)). The constraint is written as
\begin{align}
 t_{\text{end}}^{ij} \ge   t_{\text{end}}^{ik} +  \tau_{kj}^i  - M(1 - y_{kj}^i)  \quad \forall i \in \mathcal{S}, \; \forall k,j \in \mathcal{T} ,
\label{eq:temporal_constraint_with_wait}
\end{align}
where   \( M \) is chosen large enough that the inequality is automatically satisfied when \( y_{kj}^i = 0 \). For active transfers   \(  y_{kj}^i = 1 \)  the constraint reduces to the exact temporal relationship.

Finally, all repair tasks must be completed before the mission deadline \( t_{\text{max}} \). This requirement is captured by the deadline constraint
\begin{align}
t_{\text{end}}^{ik} \leq t_{\text{max}} , \quad \forall i \in S, \; \forall k \in T.
\label{eq:deadline_constraint}
\end{align}
Together, constraints \eqref{eq:temporal_constraint_with_wait} and \eqref{eq:deadline_constraint} guarantee that the mission schedule respects both the orbital-dynamics-dictated transfer times and the overall time limit for the servicing campaign.

\subsection{Optimization Problem}

The orbital routing problem can now be formulated as the following mixed-integer
optimization problem:

\begin{equation}
\min_{\mu_i,\,x_{ij},\,n_{kj}^i}
\sum_{i \in \mathcal{S}} \sum_{(k,j) \in \mu_i}
\Delta V_{kj}^i 
\end{equation}
subject to: \addtocounter{equation}{-1}
\begin{subequations} 
\begin{align}
&\sum_{i \in \mathcal{S}} x_{ij} = 1
&& \forall j \in \mathcal{T}, \\
&x_{ij} \in \{0,1\}
&& \forall i \in \mathcal{S}, \forall j \in \mathcal{T}, \\
&n_{kj}^i \in \mathbb{Z}^+
&& \forall i \in \mathcal{S}, \forall (k,j) \in \mu_i, \\
&\Delta V_i(\mu_i) \le \Delta V_{\max}
&& \forall i \in \mathcal{S}, \\
& t_{\text{end}}^{ij} \ge   t_{\text{end}}^{ik} +  \tau_{kj}^i  - M(1 - y_{kj}^i) &&  \forall i \in \mathcal{S},   \forall k,j \in \mathcal{T} \\
&t_{\text{end}}^{ik} \leq t_{\text{max}} , && \forall i \in S, \; \forall k \in T .
\end{align}
\end{subequations}

This formulation captures the coupled nature of  route selection and maneuver planning, where the ordering of tasks and the number of phasing revolutions jointly determine the orbital transfer cost.

\subsection{Discussion}

The resulting problem is a  highly combinatorial, nonconvex multi-TSP with maneuver-dependent edge costs, rendering exact solution approaches computationally intractable for realistic GEO servicing scenarios. This motivates the use of evolutionary optimization techniques, such as the proposed GA with the Route--Phasing--Split Triplet Chromosome, to efficiently compute high-quality feasible solutions.

\section{Route–Phasing–Split-Encoded Genetic Algorithm  }
\label{sec:ga_methodology}

Planning on-orbit satellite repair problem formulated in \ref{sec:problem_formulation} is a complex and highly constrained problem. Maneuver costs are nonlinear and depend on prior trajectory decisions, task execution windows are tight, and servicing multiple satellites requires maintaining both route feasibility with bounded propellant and mission time. Together, these factors create a large, combinatorial solution space that is highly sensitive to timing and orbital dynamics, making deterministic approaches prone to infeasible or suboptimal solutions.

To tackle these challenges, we employ a Route–Phasing–Split (RPS) encoded Genetic Algorithm (GA) framework that jointly explores route assignments, task sequences, and phasing decisions. Its adaptive, evolutionary search efficiently navigates the constrained, nonconvex solution space, balancing maneuver cost, feasibility, and robustness against temporal and orbital uncertainties. Chromosomes evolve through split-aware operators and feasibility-biased repair, while elite RPS solutions are further refined via a regret-based large-neighborhood search (LNS) using 2-regret destroy-and-repair and rotation polishing. This approach generates temporally coherent, low-$\Delta V$ multi-servicer plans, making it well-suited for multi-satellite repair missions.

\subsection{Chromosome Representation}
\label{subsec:chromosome_representation}

\begin{figure*}[!bp]
\centering 
\includegraphics[width=0.9\textwidth]{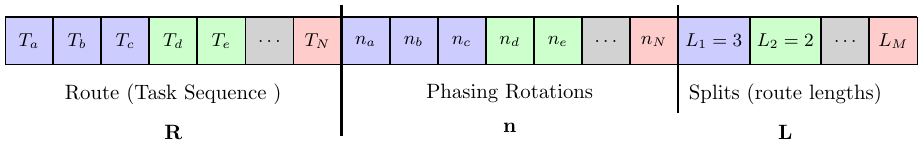}
\caption{Structure of the Route--Phasing--Split (RPS) chromosome.  Routing genes encode task order, phasing genes specify transfer rotations, and split genes define route segmentation.}
\label{fig:rps_chromosome}
\end{figure*}

Earlier, \cite{larranaga1999genetic} established permutation-based path encodings for the single-agent TSP, later extended in \cite{carter2006new} to the mTSP via composite chromosomes that decouple task allocation from subtour sequencing. GARST \cite{john2024genetic} further extended this approach for multi-agent routing with non-stationary cost. Motivated by this progression, the proposed RPS chromosome introduces a third Phasing dimension alongside Route sequencing and Split allocation, explicitly incorporating orbital synchronization requirements and enabling a unified encoding tailored to on-orbit servicing missions.

Each candidate solution is encoded using a  Route--Phasing--Split (RPS) encoded
Chromosome, designed to represent complete routing solutions while preserving feasibility.
Each chromosome $\mathbf{C}$ consists of three distinct components:
%\begin{equation}
%\mathbf{C} =
%[\underbrace{T_1, T_2, \dots, T_N}_{\text{Task Sequence}} \mid
%\underbrace{n_1, n_2, \dots, n_N}_{\text{Phasing Rotations}} \mid
%\underbrace{L_1, L_2, \dots, L_M}_{\text{Route Lengths}}]
%\end{equation}

\begin{itemize}
    \item $\mathbf{R} \  \text{(Route)} = [T_a, T_b, \dots, T_N]$ is a permutation of $N$ tasks defining the
    global execution order.
    \item $\mathbf{n} \ \text{(Phasing)}  = [n_a, n_b, \dots, n_N]$ specifies the number of phasing rotations
    $n_i \in \mathbb{Z}^+$ associated with each task transition.
    \item $\mathbf{L} \ \text{(Splitlength)}  = [L_1, L_2, \dots, L_M]$ defines the segmentation of the task sequence
    into $M$ distinct routes, with $\sum_{k=1}^M L_k = N$.
\end{itemize}

The route-length vector $\mathbf{L}$ determines how the global task sequence is partitioned
among SSCs. This triplet encoding enables the GA to simultaneously optimize task
ordering, phasing decisions, and route allocation, while ensuring that each task is executed
exactly once and all routes remain feasible by construction.

\subsection{Fitness Function Design}
\label{subsec:fitness_function}

The fitness of each chromosome is evaluated based on the total orbital maneuvering cost required to execute all assigned servicing tasks, while penalizing violations of mission-level feasibility constraints inherent to on-orbit servicing operations. In the proposed framework, minimizing propellant consumption is the primary objective, as the total required $\Delta V$ directly determines fuel usage, mission lifetime, and the number of GEO satellites that can be serviced by the available fleet.

We first define penalty terms associated with constraint violations for each servicing spacecraft.

\textit{Penalty for propellant limit violation:}
\begin{equation}
P_{\Delta V_i} = \max\!\left(0,\, \Delta V_i(\mu_i) - \Delta V_{\max}\right),
\end{equation}
 
\textit{Penalty for mission time violation:}
\begin{equation}
P_{t_i} = \max\!\left(0,\, t_{\text{end}}^{i}(\mu_i) - t_{\max}\right),
\end{equation}
where $t_{\text{end}}^{i}(\mu_i)$ denotes the completion time of the servicing route $\mu_i$, and $t_{\max}$ is the mission deadline.

The total penalty associated with SSC $S_i$ is defined as the sum of its individual constraint violations:
\begin{equation}
P_{S_i} = P_{\Delta V_i} + P_{t_i}.
\end{equation}

To penalize infeasibilities at the mission level while avoiding disproportionate dominance by any single SSC, the aggregate penalty term is defined as
\begin{equation}
P_{\text{team}} =
\left( \sum_{i \in \mathcal{S}} P_{S_i} \right)^2
+
\lambda \sum_{i \in \mathcal{S}} P_{S_i}^2.
\end{equation}
where $\lambda$ is a scaling factor.

The first term penalizes the overall magnitude of constraint violations across all the SSCs, while the second term discourages solutions in which a single SSC absorbs a large violation while others remain feasible. This combined structure mitigates oscillatory behavior in which the optimization alternates infeasibility among different SSCs without converging to a globally feasible solution.

\textit{ infeasibility penalty} A binary infeasibility penalty is introduced to penalize solutions that violate hard operational constraints. 
\begin{equation}
P_{\text{infeasible}} = \begin{cases}
\kappa , &    P_{\text{team}} \neq 0\\
0, &      P_{\text{team}} = 0  
\end{cases}    
\end{equation}
where, $\kappa $ is a large value, in this paper we choose $\kappa   = \Delta V_{\max} $ 

An infeasibility penalty is necessary because the primary mission objective is to minimize total $\Delta V$ subject to bounded propellant and mission duration constraints, rather than to minimize mission time itself. In many scenarios, a reduction in $\Delta V$ can be achieved by violating temporal constraints (e.g., through extended coasting or phasing), which may cause an otherwise feasible solution to evolve toward an infeasible one. The binary infeasibility penalty prevents such solutions from being favored during selection and ensures convergence toward physically realizable on-orbit servicing plans.

The fitness function is defined as:
\begin{equation}
\begin{aligned}
\mathcal{F} =
& \sum_{i \in \mathcal{S}} \Delta V_i(\mu_i)  
+  P_{\text{team}} +  P_{\text{infeasible}}    
\end{aligned}
\end{equation}
$\Delta V_i(\mu_i)$  represents the true physical cost of orbital transfers and is therefore the dominant component of the fitness function. The second term penalizes violations of the mission constraint. The third term is added to avoid the evolution of the feasible chromosome to an infeasible chromosome.

\subsection{Genetic Operators}
\label{subsec:genetic_operators}
All genetic operators act on the Route--Phasing--Split (RPS) triplet chromosome
\(\mathbf{C}=(\mathbf{R},\mathbf{n},\mathbf{L})\). Wherever possible operators are
component-wise (operate only on \(\mathbf{R}\), \(\mathbf{n}\) or \(\mathbf{L}\)); when
cross-component modification is required, operators are split-aware and preserve route
integrity and phasing consistency to the greatest extent possible. A lightweight repair
stage follows crossover and mutation to restore task-uniqueness and to attempt local
feasibility fixes prior to fitness evaluation.

\subsubsection{Selection strategy}

At each generation, the population is first sorted in ascending order of fitness value, where lower fitness corresponds to lower total mission cost and thus higher solution quality. An elitist selection mechanism is employed to ensure the preservation of high-quality solutions across generations. 
Specifically, the two best-performing chromosomes are copied unchanged into the next generation, thereby guaranteeing retention of the current best solutions.  
Any additional elite chromosomes, if present, are randomly permuted before insertion into the next generation. 
This permutation avoids positional bias during subsequent crossover operations while maintaining elite solution quality.

The remaining population slots are filled using roulette-wheel selection applied to an inverted fitness measure, which is appropriate for the present minimization problem.  
For each candidate chromosome $i$, the sampling weight is computed as
\begin{equation}
w_i = \max_j(f_j) - f_i + \varepsilon, \nonumber
\end{equation}
where $f_i$ denotes the total fitness value of chromosome $i$, including maneuver cost and penalty terms, and $\varepsilon = 0.01$ is a small constant introduced to prevent zero selection probability.  
The inverted formulation ensures that chromosomes with lower cost and fewer constraint violations are more likely to be selected.

The normalized selection probability is then computed as
\begin{equation}
p_i = \frac{w_i}{\sum_j w_j},  \nonumber
\end{equation}
after which a cumulative probability distribution is formed.  
Stochastic sampling with replacement is performed to generate the roulette-selected chromosomes, allowing highly fit individuals to be selected multiple times while maintaining population diversity.

The final mating pool is constructed as the concatenation of the elite subset and the roulette-wheel–selected individuals.  
This hybrid selection strategy balances exploitation of high-quality orbital routing solutions with exploration of the broader search space, which is critical for navigating the highly nonconvex and combinatorial landscape of the on-orbit servicing problem.

\subsubsection{Split-aware crossover operators}

Standard permutation-based crossover operators are inadequate for the on-orbit servicing problem, as they frequently violate route feasibility, invalidate task-phasing associations, and generate infeasible offspring. 
To overcome these limitations, a split-aware crossover family is developed that treats route blocks, as defined by the route-length vector \(\mathbf{L}\), as first-class structural elements. 
The crossover process is implemented within an adaptive, evaluate-and-select framework that explicitly prioritizes route integrity and task--phasing coherence.

\paragraph{Route-Block Crossover (RBC)}
Route-Block Crossover (RBC) exchanges entire route blocks between two parent chromosomes while preserving intra-block task order and the associated phasing information. 
The implementation proceeds as follows:
\begin{enumerate}
  \item Decompose parents \(P^A=(\mathbf{R}^A,\mathbf{n}^A,\mathbf{L}^A)\) and
  \(P^B=(\mathbf{R}^B,\mathbf{n}^B,\mathbf{L}^B)\) into route blocks according to
  \(\mathbf{L}\).
  \item Select one or more route blocks (contiguous or non-overlapping) from each parent
  using a block-selection policy (uniform, length-weighted, or improvement-guided).
  \item Exchange the selected blocks and concatenate the remaining blocks to form offspring
  task sequences; remove duplicated tasks and reinsert any missing tasks while preserving
  the original intra-block order.
  \item Inherit the phasing vector \(\mathbf{n}\) for transferred blocks from the donor
  parent and apply small, localized integer perturbations to restore
  or improve feasibility where necessary.
\end{enumerate}
By operating at the route-block level, RBC strongly preserves route-level feasibility and limits large combinatorial disruption of otherwise promising schedules.

\paragraph{Hybrid Order-Preserving Crossover (HOPC)}
When finer-grained intra-route exploration is beneficial, Hybrid Order-Preserving Crossover (HOPC) is employed. HOPC applies an order-preserving crossover restricted to tasks within selected routes, thereby preserving the relative task order inherited from donor parents while allowing limited intra-route recombination.   This operator is applied selectively, for example on short routes or when operator-credit history indicates consistent local improvements.

\paragraph{Adaptive Crossover Selection}
Both crossover operator choice and application probability are adapted online based on population fitness statistics.  
For each parent pair, the recombination probability \(p_c\) is computed using the
pair-wise best (lower) fitness \(f_i\), the population mean fitness \(\bar f\), and the
global best fitness \(f^*\):
\[
p_c =
\begin{cases}
p_{c,\max}, & f_i \le \bar f,\\[4pt]
p_{c,\max} - (p_{c,\max}-p_{c,\min})
\dfrac{f_i-\bar f}{f^*-\bar f+\varepsilon}, & f_i>\bar f,
\end{cases}
\]
This formulation ensures that higher-quality parent pairs are preferentially recombined, while weaker pairs are preserved to maintain population diversity.
A small probability of randomized location reseeding (\(p_{reseed}\)) is introduced to prevent stagnation when parent chromosomes are identical or nearly identical.

Operator selection among RBC, multi-block RBC, HOPC, and other variants is governed by a
credit-assignment mechanism. When operator \(k\) produces offspring that improve fitness
or restore feasibility, it receives a reward \(r_k\). Operator quality estimates are
updated using exponential smoothing:
\[
Q_k \leftarrow (1-\beta)Q_k + \beta r_k,
\qquad q_k = \frac{Q_k}{\sum_l Q_l},
\]
where \(\beta\) is a small learning rate controlling adaptation speed, and
\(q_k\) determines the future selection probability of operator \(k\).

\paragraph{Evaluate-and-Select Policy}
Each mating operation generates a small candidate pool by combining multiple location crossovers (OX variants), rotation realignments, and split recombinations. Rotation counts are remapped to the offspring task order, followed by mask-based mixing and small integer perturbations, and then rounded and clipped to admissible bounds.   Route-length vectors are recombined using a disruptive arithmetic split crossover to propose alternative task-to-route partitions.

Candidates are deduplicated and immediately evaluated using the GA fitness function, and only the best one or two evaluated offspring are retained to replace the parent pair.  Although this evaluate-and-select strategy increases per-mating computational cost, it significantly reduces the propagation of infeasible or high-cost offspring. This trade-off is particularly advantageous for constrained multi-TSP GEO routing problems, where maintaining route integrity and task--phasing coherence is essential for producing physically realizable servicing schedules.

\subsubsection{Mutation Strategy}
\label{subsec:mutation_strategy}

Mutation operators are applied independently to each component of the RPS triplet chromosome
\(\mathbf{C}=(\mathbf{R},\mathbf{n},\mathbf{L})\)
after preserving elites. 

Mutation rates are adaptive and fitness-dependent. For a feasible individual i we normalize
its fitness to \([0,1]\),
\[
\tilde f_i=\frac{f_i-f_{\min}}{f_{\max}-f_{\min}},
\]
and set the per-individual mutation probability
\[
p_m = p_{m,\min} + (p_{m,\max}-p_{m,\min})\,\tilde f_i,
\]
so that better solutions receive lower mutation rates. Typical implementation values are \(p_{m,\min}=0.01\) and \(p_{m,\max}=0.2\). Infeasible chromosomes are assigned a higher mutation probability to
encourage escape from infeasible basins. Component-wise mutation events are sampled
independently using the computed \(p_m\).

Route, number of phasing orbits and split length mutations are implemented as follows.

\paragraph{Route   Mutation}
- Swap Mutation: exchange two randomly selected tasks  
- Inversion Mutation: reverse a contiguous subsequence.  
- Scramble Mutation: randomly permute tasks within a chosen subsequence.  

  Mutation magnitude are annealed over generations to favor exploration early and fine-tuning later.

\paragraph{Phasing Rotation Mutation}
Phasing rotations are integer-valued and perturbed discretely. In the general model:
\[
n'_i = \max\!\big(1,\; n_i + \Delta n\cdot\operatorname{sign}(\mathcal{N}(0,\sigma_n^2))\big),
\]
where \(\Delta n\) is typically 1–2 and \(\sigma_n\) is adaptively reduced with convergence.
In the code, rotations are first remapped to the (possibly mutated) route by copying
each task's phasing rotation, an optional small integer perturbation is applied, then rotations are repaired and clipped to admissible bounds \([0,\text{max\_rot}]\). This preserves task–rotation association while enabling controlled
phasing exploration.

\paragraph{Split Length Mutation}
Operators on the split vector \(\mathbf{L}\) adjust route partitioning with conservative,
local moves:
\begin{itemize}

\item Route Splitting: divide a long route into two at a chosen boundary to reduce local infeasibility or cost.

\item Route Merging: combine adjacent routes when the merged route can be feasibly repaired or yields net benefit. 

\item Boundary Adjustment (Transfer): move a small number of tasks between two routes. It typically transfers 1–2 tasks between randomly selected routes.  
\end{itemize}
All split mutations are repaired to preserve chromosome's structure.

\paragraph{Repair and Practical Notes}
Every mutation step is followed by inexpensive, targeted repairs (task-uniqueness enforcement, rotation repair, and split-sum repair) before subsequent evaluation. This adaptive, feasibility-biased mutation strategy concentrates fine-grained search in promising regions while aggressively perturbing infeasible or poorly performing chromosomes, balancing exploration and exploitation in the constrained multi‑TSP GEO routing problem.

\subsection{Large-Neighborhood Search (LNS) Intensification}
\label{subsec:lns}

To enhance the GA’s global exploration while improving local solutions, we embed a Large-Neighborhood Search (LNS) module as a local improvement module applied to the RPS triplet chromosome. One of the elite triplets, encoding task routes \(\mathbf{R}\), phasing rotations \(\mathbf{n}\), and route lengths \(\mathbf{L}\), serves as the LNS seed. The method repeatedly performs destroy-and-repair moves on this seed, enabling substantial, structured adjustments across ordering, phasing and partitioning that crossover or mutation alone often cannot achieve. LNS returns the best solution found within a fixed iteration budget and is applied selectively to concentrate computational effort on the most promising regions of the search space. Each iteration proceeds through three steps: destroy, repair and accept.

\paragraph{Destroy}  
Destruction generates a partial RPS triplet by removing a fraction of tasks (defined by a destruction rate \(\rho\)) from the task sequence while maintaining route-split and phasing structure. To balance exploration and targeted refinement, three strategies are rotated: (i) random deletion, (ii) high-cost deletion, removing tasks that contribute disproportionately to route-level \(\Delta V\), and (iii) cluster deletion, removing a contiguous subsequence from the longest route. These strategies trade broad exploration (random) for impact-focused restructuring (high-cost) and coherent local reallocation (cluster).

\paragraph{Repair}  
Repair reconstructs a feasible RPS triplet using a regret-based, time-aware insertion heuristic. For each unassigned task, feasible insertion positions across servicers and route segments are evaluated, accounting for waiting, coasting, phasing rotations, and repair durations. Denoting the best and second-best insertion costs by \(c^{(1)}\) and \(c^{(2)}\), the 2-regret value is
\[
\mathrm{regret}(t) = c^{(2)}(t) - c^{(1)}(t).
\]
Tasks with maximum regret are inserted first at their optimal positions, reusing stored phasing rotations where possible and updating the route-length vector \(\mathbf{L}\) accordingly. This approach mitigates myopic placements and produces temporally coherent, feasible reconstructions.

\paragraph{Accept}  
Repaired triplets are evaluated with the GA’s penalized fitness. Acceptance may be greedy or probabilistic: non‑improving candidates can be accepted with probability
\[
P_{\text{accept}}=\exp\!\big(-\Delta f / T\big),\qquad \Delta f=f_{\text{new}}-f_{\text{current}}>0,
\]
where \(T\) is a temperature parameter. Best-so-far bookkeeping preserves the best feasible (or least‑penalized) solution across the iteration budget.

\paragraph{Rotation neighbourhood search (phasing polish)}  
In addition to destroy-and-repair LNS, we apply a lightweight rotation neighbourhood search that perturbs only the integer phasing vector \(\mathbf{n}\) while holding \(\mathbf{R}\) and \(\mathbf{L}\) fixed. The neighbourhood consists of single‑unit adjustments
\[
\mathbf{n}' = \mathbf{n} + \boldsymbol{\delta}, 
\qquad 
\boldsymbol{\delta} \in \{-1,0,1\}^k, 
\qquad 
\mathbf{1}^\top \boldsymbol{\delta} = 0,
\]

(clipped to admissible bounds \(1\le n_i\le R_{\max}\)) and a permutation‑augmented copy of the same single‑step moves to permit nonlocal reassignment of phasing patterns.  Because small changes in \(\mathbf{n}\) often produce large, interpretable effects on transfer geometry and schedule feasibility, the rotation search provides high‑leverage, low‑cost improvements. 

\paragraph{Remarks}  LNS is applied selectively, typically once per generation or when convergence stagnates, to limit computational overhead. By exploring small, feasible perturbations around the current best RPS triplet chromosome, it complements the GA’s global search, refining task ordering, phasing, and route assignments without disrupting population diversity. This synergy between global exploration and targeted local exploitation improves the likelihood of identifying low-$\Delta V$ feasible routes while preserving task–phasing and route integrity, particularly in the highly constrained multi-TSP GEO servicing problem.

\subsection{Stagnation Detection, Diversity Injection, and Convergence}\label{subsec:stagnation_diversity_convergence}
On-orbit routing is highly combinatorial and constrained, producing a search landscape with numerous deep local minima and narrow feasible regions; without intervention, the population can rapidly collapse around a single incumbent, eliminating the exploratory variability needed to discover alternative RPS configurations. A stagnation-control mechanism is therefore required, not to micromanage the search, but to restore the capacity to explore structurally different solutions while preserving elites. 

\paragraph{Stagnation Detection} 
To mitigate premature convergence, we employ a dual-criterion stagnation detection mechanism and inject controlled diversity when necessary. After an elitist merge of parents and offspring, duplicate genotypes (identical RPS chromosomes), are removed while preserving the relative ordering of remaining individuals. The population is then sorted by ascending fitness.

To monitor convergence, the population is considered compressed if the fitness of a near-tail individual (e.g., the 90\% percentile) is within $0.5\%$ of the best fitness, indicating that exploration has slowed.

Let \(f_{(1)}^{(t)}\) denote the best fitness at generation \(t\); stagnation is declared only if the population is compressed and the relative improvement of the best fitness over a window of \(w\) generations satisfies
\[
\frac{|f_{(1)}^{(t)} - f_{(1)}^{(t-w)}|}{f_{(1)}^{(t-w)}} < \delta.
\]
In this work, we select a window of \(w = 5\) generations and a tolerance of \(\delta = 10^{-4}\).

\paragraph{Diversity Injection}
Upon detecting stagnation, a substantial diversity injection is applied, replacing a large fraction (e.g., 90\%) of the population with new individuals drawn from deeper entries in the merged, deduplicated pool:
\[
\mathcal{P}_{N-b+1:N} \leftarrow \mathcal{P}^{\text{merged}}_{j:j+b-1}, \quad b \approx 0.9 N, \quad j > N,
\]
where \(\mathcal{P}^{\text{merged}}\) is the merged, sorted, and deduplicated population. Unlike random reinitialization, this approach preserves structural features of the current search context, producing individuals that are more amenable to subsequent repair and local improvement.

\paragraph{Convergence}
Finally, convergence is ensured through a stopping criterion based on the improvement of the best fitness. If the relative improvement over the last 50 generations satisfies
\[
\frac{|f_{(1)}^{(t)} - f_{(1)}^{(t-50)}|}{f_{(1)}^{(t)}} < \delta,
\]
the algorithm is terminated. This prevents unnecessary computation once the search has effectively stabilized while ensuring sufficient opportunity for exploration, particularly in complex, multi-target routing scenarios.

The algorithm delays diversity injection until after burn‑in, measures compression using a near‑tail metric, removes duplicate genotypes, and replaces individuals with contextually related ones rather than random candidates. These choices reintroduce useful diversity without losing valuable learned structure. They preserve elite solutions, give tunable controls (replacement fraction, compression threshold, stagnation window), and help the algorithm escape stagnation to converge on feasible, low $\Delta V $  solution.

\subsection{Discussion}
\label{subsec:ga_discussion}

The proposed genetic algorithm evolves the population through selection, split-aware crossover, adaptive mutation, and local neighborhood search. To prevent premature convergence, stagnation is detected via population compression and relative fitness improvement, triggering controlled diversity injection of structurally distinct, contextually related individuals. The algorithm stops when either a maximum generation limit is reached or improvement over a defined window falls below a tolerance.

This integrated framework efficiently converges to feasible, low-$\Delta V$ on-orbit routing solutions while respecting multi‑TSP GEO constraints. By combining elitist preservation, forward-invariant evolution, targeted diversity injection, and local refinement, it balances exploration and exploitation for robust multi-target mission planning. The resulting performance is presented in the following section.

\section{Results} \label{sec:results}

 All algorithms were implemented in the MATLAB 2024b environment and executed on a PC equipped with a 1.8 GHz Intel Core i7 CPU and 48 GB RAM.

\subsection{Case Study} 
To evaluate the performance of the proposed method, a real-world mission scenario reported in \cite{han2022multiple} is considered. All algorithms were developed in MATLAB R2024b and executed on a workstation equipped with a 1.8 GHz Intel Core i7 processor and 48 GB of RAM to ensure a consistent computational environment for performance assessment.
Table \ref{tab:orbit_params} lists the orbital parameters of 2 SSCs and 14 target satellites at the commencement of the mission. Each SSC is constrained by a maximum total $\Delta V$ budget of 1000 m/s. All targets must be serviced within a 30-day mission horizon, and the on-orbit servicing duration for each satellite is fixed at 20 hours. The parameters for RPS based GA  are set as follows: the minimum iteration number is set to 100, the population size is set to 100, the  the crossover probability adaptively changes from 0.6 to 0.9, the mutation probability adaptively changes from 0.08 to 0.2, the remove rate for Destroy method is 30\%, the iteration number of LNS procedure is fixed to 5. The infeasibility penalty coefficient $\kappa$ is 1000. The stopping criteria comprise two conditions; the total number of iterations must exceed the prescribed minimum threshold, and  the best fitness value must remain unchanged over the most recent 50 consecutive iterations, indicating convergence. To evaluate robustness and statistical stability, the mission scenario is independently solved 100 times using the proposed RPS-based GA algorithm.

\begin{table}[tbhp!]
\centering
\caption{Orbital Parameters of Servicing Spacecraft (SSC) and Target Satellites}
\label{tab:orbit_params}
\begin{tabular}{llrrr}
\toprule
\textbf{ID} & \textbf{Name} & \textbf{\begin{tabular}[c]{@{}c@{}}Inclination \\ ($i^\circ$)  \end{tabular}}  & \textbf{\begin{tabular}[c]{@{}c@{}} \qquad RAAN \\ \qquad  ($\Omega^\circ$)  \end{tabular}}    &  \textbf{\begin{tabular}[c]{@{}c@{}}True Anomaly  \\  ($\omega^\circ$)   \end{tabular}}    \\
\midrule
SSC1 & SSC1 & 0.00 & 0.00 & 0.00 \\
SSC2 & SSC2 & 5.00 & 0.00 & 160.00 \\
$T_1$ & Beidou2\_G7 & 1.60 & 66.76 & 278.27 \\
$T_2$ & Beidou2\_G8 & 0.30 & 328.08 & 156.03 \\
$T_3$ & Beidou\_G1 & 1.80 & 45.11 & 252.16 \\
$T_4$ & Beidou\_G2 & 7.77 & 52.63 & 328.00 \\
$T_5$ & Beidou\_G3 & 1.89 & 52.10 & 274.21 \\
$T_6$ & Beidou\_G4 & 1.06 & 59.65 & 144.68 \\
$T_7$& Beidou\_G5 & 1.45 & 67.40 & 288.52 \\
$T_8$ & Beidou\_G6 & 1.86 & 85.65 & 319.30 \\
$T_9$ & Chinasat\_11 & 0.09 & 103.25 & 331.94 \\
$T_{10}$ & Fengyun\_2E & 5.00 & 68.04 & 285.07 \\
$T_{11}$ & Fengyun\_2F & 2.80 & 83.11 & 224.48 \\
$T_{12}$& Tianlian1\_01 & 4.81 & 71.74 & 337.75 \\
$T_{13}$ & Tianlian1\_02 & 2.21 & 74.98 & 229.24 \\
$T_{14}$ & Tianlian1\_03 & 0.99 & 98.18 & 230.86 \\
\bottomrule
\end{tabular}
\end{table}

\begin{figure}[htbp!]
        \centering
        \includegraphics[width=\linewidth]{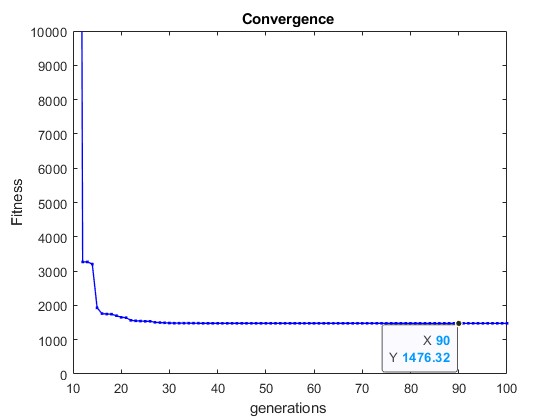}
        \caption{Convergence Curve for RPS-based LNSAGA}
        \label{fig:convergence}
\end{figure}

Figure \ref{fig:convergence} shows the convergence curve for the best run among 100 trials. The algorithm converges to an optimal total $\Delta V$ of 1476.32 m/s, significantly lower than the previously reported in \cite{han2022multiple}  best values of $1956.36$ m/s for LNS-AGA and $2285.59$ m/s for standard GA. 

\begin{figure}[htbp!]
        \centering
        \includegraphics[width=\linewidth]{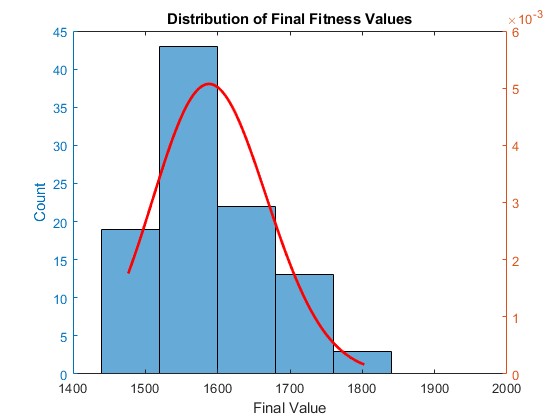}
        \caption{Distribution of the final values obtained for randomised simulations}
        \label{fig:distribution}
\end{figure}

\begin{figure*}[hbt!]
    \centering
    % --- First subfigure ---
    \begin{subfigure}[b]{0.45\textwidth}  % 45% of text width
        \centering
        \includegraphics[width=\textwidth]{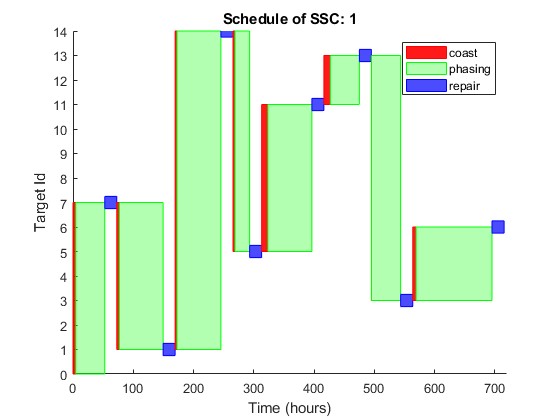}
        \caption{ Repairing schedule of SSC 1}
        \label{fig:schdeule_1}
    \end{subfigure}
    \ 
    % --- Second subfigure ---
    \begin{subfigure}[b]{0.45\textwidth}
        \centering
        \includegraphics[width=\textwidth]{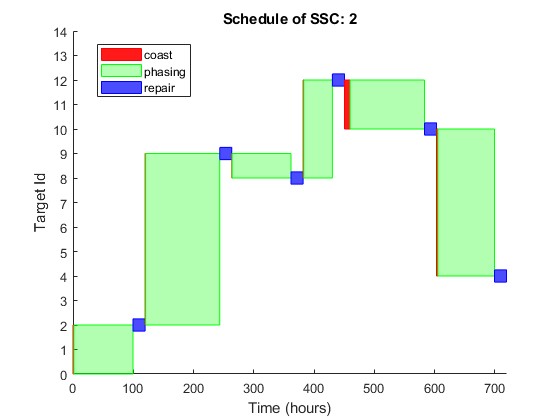}
        \caption{Repairing schedule of SSC 2}
        \label{fig:schdeule_2}
    \end{subfigure}
    
    \caption{Schedules of the SSC for entire mission}
    \label{fig:schdeule}
\end{figure*}

Figure \ref{fig:distribution} reports the histogram of final fitness values for 100 runs. All of the runs results in a feasible solution, and final fitness values follow the gamma distribution as shown in Fig \ref{fig:distribution} the histogram. 
Fig \ref{fig:schdeule}, shows the schedule of both SSCs for the optimal solution. One can observe that  all the targets have been repaired. All tasks in the on-orbit repair mission are completed within the mission deadline of 720 hours, i.e., 30 days.   Unlike the results reported in \cite{han2022multiple}, the schedule shows that the phasing time is dynamic and significantly varies due to dynamic number of phasing rotations.

Table \ref{tab:mission_results} presents the detailed mission schedule, including the start time of each task, coast time, first impulse vector, phasing time, number of phasing revolutions, second impulse vector, task completion time, and the corresponding $\Delta V$ magnitude. The time of the first impulse, ${\Delta v}{1}$, is computed as the sum of the task's start time and the coast time. The second impulse, ${\Delta v}{2}$, occurs at the sum of the task's start time, coast time, and phasing time. After rendezvous, the SSC performs on-orbit servicing for 20 hours, upon which the task is completed. From Table \ref{tab:mission_results}, SSC1 consumes a total $\Delta V$ of 586.09 m/s, while SSC2 consumes 890.23 m/s. Both servicing spacecraft operate within the maximum allowable individual SSC's $\Delta V$ limit of 1000 m/s. Both SSCs complete their respective servicing tasks within the temporal constraint of 720 hours.

{\renewcommand{\arraystretch}{1.3}
\begin{table}[htb!]
\centering
\caption{Detailed mission planning results for  orbital routing details }
\label{tab:mission_results}
\resizebox{\textwidth}{!}{%
\begin{tabular}{ccrrcrccrr}  \hline 
Id &
  Name &
  \multicolumn{1}{c}{\begin{tabular}[c]{@{}c@{}}start\\ time (h)\end{tabular}} &
  \multicolumn{1}{c}{\begin{tabular}[c]{@{}c@{}}coast\\   time (h)\end{tabular}} &
 $\mathbf{\Delta v}_{\text{1}}$  &
  \multicolumn{1}{c}{\begin{tabular}[c]{@{}c@{}}phasing\\ time (h)\end{tabular}} &
  \multicolumn{1}{c}{\begin{tabular}[c]{@{}c@{}}Phasing \\ rotations\end{tabular}} &
   $\mathbf{\Delta v}_{\text{2}}$ &
  \multicolumn{1}{c}{\begin{tabular}[c]{@{}c@{}}completion\\time(h) \end{tabular}} &
  \multicolumn{1}{c}{ ${\Delta v}_{\text{total}} $} \\   \hline 
SSC1  &    &    &         &       &    &         &   &    &     \\
$T_7$  & Beidou\_G5    & 0.00   & 4.48  & [-4.42,1.84,77.80 ]   & 48.14  & 2 & [5.33,-2.22,0.00]  & 72.62  & 83.73  \\
$T_1$  & Beidou2\_G7   & 72.62  & 4.28  & [-8.71,4.94,8.36]   & 72.53 & 
3 & [8.91,-5.02,-0.26]     & 169.43 & 23.27  \\
$T_{14}$ & Tianlian1\_03 & 169.43 & 3.10  & [-9.06,12.83,-48.73]  & 72.87 & 3 & [8.01,-12.62,-0.34]    & 265.40 & 66.15  \\
$T_5$  & Beidou\_G3    & 265.40 & 3.41  & [-1.20,6.51,75.06]     & 24.12 & 1 & [2.82,-7.18,-0.03]   & 312.92 & 83.07  \\
$T_{11}$ & Fengyun\_2F   & 312.92 & 10.30 & [-11.65,-10.49,82.21]   & 73.05 & 3 & [14.71,9.41,-0.19]     & 416.28 & 101.16 \\
$T_{13}$  & Tianlian1\_02 & 416.28 & 10.69 & [-6.06,-1.35,-36.64]  & 48.09  & 2 & [4.48,1.65,-0.21 ]     & 495.06 & 41.94  \\
$T_3$   & Beidou\_G1    & 495.06 & 0.19  & [9.37,5.22,59.13]   & 48.33 & 2 & [-7.60,-6.18,0.22 ]   & 563.58 & 69.89  \\
$T_6$  & Beidou\_G4    & 563.58 & 5.75  & [-22.99,45.86,-42.41] &125.85 & 5 & [22.15, -45.17, -1.49] &715.18 & 116.89    \\   \noalign{\vspace{1mm}} % vertical space before the line
\cline{9-10}
\noalign{\vspace{1mm}} % vertical space after the line
 &  &  &  &  &  &  &   &  {\textbf{715.18}}  & \multicolumn{1}{r }{\textbf{586.09}} \\  
 \hline
   SSC2    &   &        &      &   &        &   &   &        &       \\  
$T_2$ & Beidou2\_G8   & 0.00   & 1.46 & [0.48,-36.49,252.28]  & 98.12  & 4 & [-0.83,24.82,2.17]  & 119.58 & 279.83 \\
$T_9$ & Chinasat\_11  & 119.58 & 0.90 & [-18.09,-20.25,19.68] & 122.92 & 5 & [18.08,20.21,0.14]  & 263.41 & 60.66  \\
$T_8$ & Beidou\_G6    & 263.41 & 0.56 & [-19.38,1.82,95.23]   & 97.75  & 4 & [20.99,-1.92,-0.03] & 381.72 & 118.28 \\
$T_{12}$  & Tianlian1\_01 & 381.72 & 1.11 & [3.32,0.58,162.91]    & 47.57  & 2 & [5.81,-2.92,-0.20]  & 450.40 & 169.45 \\
$T_{10}$ & Fengyun\_2E   & 450.40 & 9.27 & [4.23,-29.97,-21.00]  & 123.42 & 5 & [-5.82,30.55,1.27]  & 603.09 & 67.97  \\
$T_{4}$   & Beidou\_G2    & 603.09 & 2.18 & [6.87,7.58,173.78]    & 93.91  & 4 & [9.47,-17.53,-1.34] & 719.19 & 194.05 \\  
  \noalign{\vspace{1mm}} % vertical space before the line
\cline{9-10}
\noalign{\vspace{1mm}} % vertical space after the line
 &  &  &  &  &  &  &  & {\textbf{719.19}} & \multicolumn{1}{r}{\textbf{890.23}} \\  
 \hline
\end{tabular} }
\end{table} }

Table \ref{tab:compare_results} presents the comparison of the mission with the state-of-the-art LNS-AGA method proposed in \cite{han2022multiple}. The proposed RPS-GA finds the solution with total $Delta V = 1476.32$, which is $24.5\%$ lower than the previously known best value ($1956.32$). Both SSCs satisfy the temporal constraint of 720 hrs. 

{\renewcommand{\arraystretch}{1.1}
\begin{table}[htbp]
\centering
\caption{Comparison table}
\label{tab:compare_results}
\begin{tabular}{|l|l|r|r|}
 \hline
 &  & $\Delta V  (m/s)$ & \makecell{Completion\\time (hr)} \\ \hline
 Constraints &   &1000 & 720\\ \hline 
\multirow{3}{*}{LNS AGA \cite{han2022multiple}} 
 & SSC1 & 992.68 & 607.45 \\ \cline{2-4}
 & SSC2 & 963.68 & 703.33 \\ \cline{2-4}
 & Mission & 1956.36 & 703.33 \\ \hline

\multirow{3}{*}{RPS-GA} 
 & SSC1 & 586.09 & 715.18 \\ \cline{2-4}
 & SSC2 & 890.23 & 719.19 \\ \cline{2-4}
 & Mission & 1476.32 & 719.19 \\ \hline

\end{tabular}
\end{table}
}
\subsection{Ablation Study}

\begin{figure*}[htb!]
    \centering
    \begin{subfigure}[b]{0.45\textwidth}
        \centering
        \includegraphics[width=\textwidth]{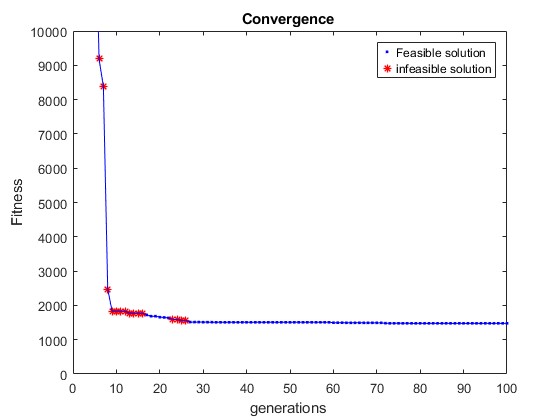}
        \caption{Feasible solution}
        \label{fig:ablation_feasible}
    \end{subfigure} \  \ 
    \begin{subfigure}[b]{0.45\textwidth}  % 45% of text width
        \centering
        \includegraphics[width=\textwidth]{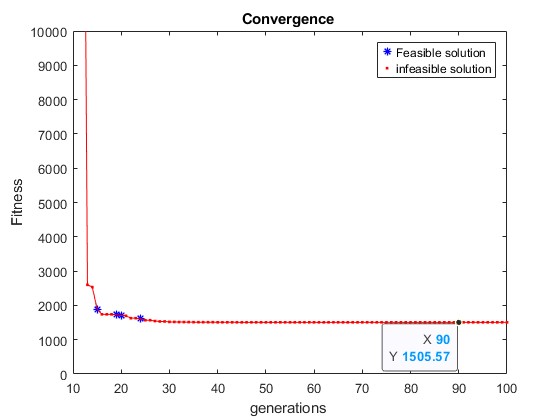}
        \caption{ Infeassible solution }
        \label{fig:ablation_infeasible}
    \end{subfigure}      
    \caption{Convergence Curves without infeasibility penalty}
    \label{fig:ablation}
\end{figure*}

The infeasibility penalty is discrete in nature and is introduced to prevent convergence toward infeasible local minima. To evaluate its necessity, the proposed RPS-based LNS-AGA was simulated with the infeasibility penalty coefficient set to zero.Out of 100 runs, 12 converged to infeasible solutions, and several runs that produced feasible solutions exhibited oscillations between feasible and infeasible states during evolution. Without the infeasibility penalty, the proposed RPS-based GA converges to the best solution with $\Delta V = 1476.32$; however, during evolutions, the solution toggles between the feasible and infeasible chromosomes as shown in the fig \ref{fig:ablation_feasible}.

One representative infeasible example (the lowest infeasible solution) is shown in Fig. \ref{fig:ablation_infeasible}, where the converged solution is infeasible. Although feasible individuals exist in the final population, the additive fitness structure allows an infeasible solution to attain the best objective value in the absence of an infeasibility penalty term. 

In this infeasible run, SSC1 is assigned the sequence   $  \{ SSC_1 $ - $T_{14}$ - $T_{5}$ - $T_{3}$ - $T_{13} $- $T_{11} $- $T_{10} $- $T_{12}$  - $T_{4} \}$ with phasing rotations (3, 1, 3, 1, 1, 4, 5, 3), resulting in a total $\Delta V$ of 816.95 m/s and a completion time of 710.94 hours. SSC2 is assigned $\{SSC2$ - $T_{6}$ - $T_{2}$ - $T_{9}$ - $T_{8}$ - $T_{7}$ - $T_{1} \}$,
with phasing rotations (4, 5, 5, 3, 5, 2), resulting in a total  $\Delta V$ of 686.91 m/s and a completion time of 720.92 hours, thereby violating the temporal constraint. 

The results show that without an infeasibility penalty, the algorithm favors infeasible solutions and oscillates between feasible and infeasible individuals, highlighting the penalty’s essential role in ensuring forward-invariant evolution within the feasible solution space.

\section{Conclusion} \label{sec:conclusion}
This paper presented a Route–Phasing–Split Genetic Algorithm (RPS-GA) for multi-servicer on-orbit servicing (OOS) mission planning in geosynchronous Earth orbit (GEO). The proposed  RPS-GA approach integrates route, phasing, and split decisions within a single optimization framework. The core element is an RPS triplet chromosome that jointly encodes route order, phasing rotations, and route partitions, enabling split-aware recombination while preserving high-quality multi-servicer route blocks. The proposed constraint-aware fitness function compares feasible solutions solely by total $\Delta V$, penalizing both aggregate and unbalanced violations of propellant and temporal constraints while preserving a forward-invariant feasibility of the best solution. The approach is further strengthened by tailored split-aware genetic operators, including Route–Block Crossover, adaptive mutation, and a regret-based Large Neighborhood Search for local improvement and exploitation of favorable phasing opportunities.
Computational experiments on representative GEO servicing scenarios demonstrate that RPS-GA produces feasible multi-servicer plans with substantially improved propellant efficiency, reducing total $\Delta V$   from 1956.36 m/s to 1476.32 m/s compared with a state-of-the-art LNS-AGA baseline, in the same scenario. These result  highlight the benefits of explicitly co-optimizing route sequencing, phasing rotations, and route partitioning, combined with split-aware evolutionary operators and robust feasibility management, for realistic multi-servicer GEO OOS planning.

%

%\appendices
%\section{}

%\section{}

\section*{Acknowledgment}
This work was partially supported by the Wallenberg AI, Autonomous Systems and Software Program (WASP) funded by the Knut and Alice Wallenberg Foundation, and partially by the OPTACOM a research project funded by European Space Agency (ESA).

\bibliography{references.bib}

\end{document}